\documentclass[conference,12pt,]{IEEEtran}
\IEEEoverridecommandlockouts
\usepackage{cite}
\usepackage{amsmath,amssymb,amsfonts}
\usepackage[utf8]{inputenc}
\usepackage{algorithmic}
\usepackage{graphicx,color}
\usepackage{textcomp}
\usepackage{cases}
\usepackage{url}

\def\BibTeX{{\rm B\kern-.05em{\sc i\kern-.025em b}\kern-.08em
    T\kern-.1667em\lower.7ex\hbox{E}\kern-.125emX}}

\usepackage{xcolor}
\newcommand{\vect}[1]{\boldsymbol{#1}}

\newcommand{\grade}[1]{\left\langle #1\right\rangle}
\author{Seil Sautbekov}
\title{Covariant Electrodynamics with a Scalar Degree of Freedom}
\begin{document}
\maketitle

\begin{abstract}
	Within the Clifford algebra of spacetime, a covariant electrodynamics is formulated in which the derivative of the four-potential contains a scalar field (S) in addition to the electromagnetic bivector. A generalized system of Maxwell equations is derived that reduces to the classical Maxwell system when S=0. The energy–momentum tensor, the corresponding conservation laws, and a generalized Lorentz force density are obtained. It is shown that, in free space, the field S admits wave solutions, termed scalar S-waves. The structure and energy characteristics of a free S-wave, as well as its possible sources and the conditions for its experimental detection, are determined.
\end{abstract}	

\section{Introduction}
\label{sec:introduction}

Classical electrodynamics is one of the most extensively
experimentally validated physical theories. Maxwell's equations
provide a unified description of electrostatic and magnetic fields,
electromagnetic induction, radiation, and the propagation of
electromagnetic waves. Polarization experiments in optics have
firmly established the transverse nature of conventional
electromagnetic radiation in free space.

At the same time, the history of electrodynamics includes a distinct
class of Tesla's experiments with high-voltage resonant systems,
including single-wire and ground-connected schemes for energy
transmission \cite{Tesla1897,Tesla1900}. The field geometry in such
systems differs substantially from that of a plane wave in free
space and involves conductive, capacitive, inductive, and
surface-wave mechanisms. Modern implementations of similar systems
can generally be described within classical electrodynamics and do
not, by themselves, demonstrate the existence of an additional
scalar mode \cite{Bolshev2023,Sergeichev2018}. However, the known
experiments were generally not designed to reconstruct the complete
spacetime structure of the field or to test a specific criterion
capable of distinguishing such a mode from conventional near-field,
guided, and surface-wave components.

Therefore, any extension of electrodynamics must satisfy the
correspondence principle: within the domain of established
phenomena, it must reproduce the standard Maxwell system and may
differ from it only under specifically defined conditions.

In a covariant formulation, the electromagnetic field is constructed
from the four-potential $A$. In spacetime algebra (STA) \cite{hestenes1966spacetime}, its full
derivative admits the following identity decomposition by grade:
\begin{equation}
 \nabla^{(4)}A
 =
 \nabla^{(4)}\!\cdot A
 +
 \nabla^{(4)}\!\wedge A .
 \label{eq:intro_full_derivative}
\end{equation}
The bivector part
\begin{equation}
 G=\nabla^{(4)}\!\wedge A
 \label{eq:intro_bivector}
\end{equation}
defines the standard electromagnetic field, whereas the scalar part
\begin{equation}
 S=\nabla^{(4)}\!\cdot A
 \label{eq:intro_scalar}
\end{equation}
is not regarded as an independently observable field in conventional
electrodynamics. Its value depends on the choice of gauge for the
potential, and the Lorenz gauge condition allows one to set $S=0$
without changing the observable electromagnetic fields
\cite{JacksonOkun2001}.

Exterior-algebra formulations have generalized Maxwell
fields of a fixed grade to space-times with arbitrary numbers
of spatial and temporal dimensions~\cite{Colombaro2020}.
In these formulations, the field is an $r$-vector defined by
the exterior derivative $F=\partial\wedge A$, whereas the source has grade $r-1$. 

The present work follows a different direction: it retains the physical $(1,3)$ spacetime but uses
the full geometric derivative $\partial A$, thereby combining the scalar and bivector parts in a single mixed-grade Clifford field.

The standard $S=0$ description is fully consistent with the vast range of experimentally investigated electromagnetic phenomena. In particular, radiation from conventional sources in free space is
transverse, whereas longitudinal electric-field components arising
in the near field, structured and tightly focused beams, guiding
systems, and material media are described by the standard Maxwell
equations \cite{Lekner2016,Maluenda2021}. Consequently, the mere
presence of a longitudinal electric field does not justify the
introduction of an additional wave mode.

Nevertheless, the experimental success of the $S=0$ sector does not
exhaust the more general question of whether the scalar part of the
full derivative of the four-potential can acquire physical content
of its own in an extended system in which the status of gauge freedom
is modified accordingly. Such a formulation is physically meaningful
only if retaining the scalar part is consistent with the established
conservation laws, yields a positive energy density, and permits
observable consequences that are absent when $S=0$.

In the present work, the scalar part of the full derivative of the
four-potential is retained as an independent degree of freedom.

The complete field is defined by the mixed-grade multivector
\begin{align}
F=S+G.\label{eq:intro_full_field}
\end{align}
In the limit $S\rightarrow0$, the extended system reduces to standard
Maxwell electrodynamics. The correspondence principle is therefore
satisfied at the level of the field equations, energy, momentum, and
the associated conservation laws. The purpose of this work is not to
revise established results of classical electrodynamics, but to
investigate the physical content of that part of the covariant
geometric field structure which is eliminated by a gauge condition
in the standard formulation.

The resulting system makes it possible to formulate a distinct
experimental problem: to determine a source configuration and a
measurement criterion under which the contribution of a free
$S$-mode can be unambiguously distinguished from longitudinal
near-field components, conduction currents, capacitive and inductive
coupling, and guided and surface waves. The present work establishes
the theoretical signatures of such a mode and the conditions
necessary for its detection; the design of a specific experimental
setup and its metrological implementation are beyond the scope of
this paper.

The paper is organized as follows. In
Sec.~\ref{sec:field_equations}, the extended system of field
equations for the $S$, $\vect E$, and $\vect B$ components is derived
from the full derivative of the four-potential, and its relation to
Maxwell's equations is established. In
Section \ref{sec:energy_momentum}, the quadratic energy--momentum
operator is constructed, and the energy density, energy flux,
momentum density, and spatial stress tensor are determined. The
corresponding local conservation laws are derived in
Section \ref{sec:conservation_laws}. Section \ref{sec:s_wave} examines free plane and
spherical $S$-waves and their energy structure. Section \ref{sourcesection} considers
the mechanism of scalar-field generation, whereas Section \ref{sec:discussion}
discusses its physical significance, possible applications, and
criteria for experimental identification. The principal results are
summarized in the Conclusion.

\section{Electromagnetic Potential}
\label{sec:field_equations}

The source of the electromagnetic field is the four-current
\begin{equation}
J=(c\rho, \vect j)
\end{equation}
in Minkowski spacetime, where $\rho$ is the charge density,
$\vect j$ is the current density in three-dimensional space, and
$c=(\varepsilon_0\mu_0)^{-1/2}$ is the speed of light.

The electromagnetic field $F$ in Minkowski spacetime is defined by
the sequence
\begin{equation}\label{Tr}
J\;\xrightarrow{\;{*}\;}\;A\xrightarrow {\;\partial\;} F,
\end{equation}
where the symbol $*$ denotes convolution over all coordinates
$x=(ct, \vect r)$, and $\partial$ is the derivative operator.

The physical quantity
\begin{equation}\label{A}
A(\vect r,t)=\big(\varphi(\vect r,t)/c, \vect A(\vect r,t)\big)=\mu_0 J*\psi(\vect r,t),
\end{equation}
traditionally referred to in electrodynamics as the retarded
potential, is defined by the convolutions
\begin{align}
\varphi(\vect r,t)=\frac1\varepsilon_0\psi(\vect r,t)*\rho(\vect r,t), \label{phi}\\
\vect A(\vect r,t)=\mu_0\,\psi(\vect r,t)*\vect j(\vect r,t) \label{Aphi}
\end{align}
with the Green function
\begin{equation}
\psi(\vect r,t)=\frac{\delta(t-r/c)}{4\pi r}.
\end{equation}

Thus, the four-potential $A$ is treated here as a physical quantity
satisfying the wave equation
\begin{equation} \label{DA}
\Box\, A(x)=\mu_0\,J(x).
\end{equation}

Indeed, Eq.~\eqref{DA} follows from Eq.~\eqref{A} by applying the
d'Alembert operator ($\Box$) and using the identity
\begin{equation*}
(\Box\,\psi)*f\equiv \delta^{(4)}(x)*f=f
\end{equation*}
in accordance with the properties of convolution.

In the present formulation, $A$ is not treated as an
arbitrary representative of a gauge-equivalence class. It is
defined as the causal four-vector field generated by the
physical source through the retarded Green function. Once
the source and the initial and boundary conditions are fixed,
this prescription determines $A$ uniquely. A transformation
$A\rightarrow A+\partial\chi$ preserves the same source
equation only if $\Box\chi=0$; the remaining homogeneous
freedom is further restricted by the imposed retarded
conditions. Moreover, for every admissible residual
transformation,
\[
 S\rightarrow S+\Box\chi=S,
 \qquad
 G\rightarrow G+\partial\wedge\partial\chi=G.
\]
Thus, the scalar component $S$ is invariant under the
residual transformations compatible with the defining
source equation and its boundary conditions.

It is evident that the three-dimensional components of the vector
$A$ also satisfy the wave equations
\begin{align}
&\Box\, \varphi(\vect r,t)=\frac1\varepsilon_0\,\rho(\vect r,t), \nonumber\\
&\Box\vect\, A(\vect r,t) =\mu_0\,\vect j(\vect r,t). \nonumber
\end{align}

It should be noted that treating the four-potential as a physical
quantity is consistent with the Aharonov--Bohm effect, according to
which the potential affects the phase of the wave function of a
charged particle in the region of its motion where the
electromagnetic field locally vanishes.

Let $A=A^\mu\gamma_\mu$ be the four-potential and
$\nabla_{(\!4)}=\gamma^\mu\partial_\mu$ be the four-dimensional
vector derivative in $\textrm{Cl}_{1,3}$ with signature
$\eta_{\mu\nu}=\eta^{\mu\nu}=\operatorname{diag}(+1,-1,-1,-1)$.

Define
\begin{equation}
I=\gamma_0\gamma_1\gamma_2\gamma_3,
\qquad I^2=-1 .
\end{equation}
The geometric product of two vectors contains scalar and bivector
parts. Therefore, in terms of the inner $(\cdot)$ and outer
$(\wedge)$ products, the derivative of the potential identically
decomposes into two components:
\begin{equation}\label{eq:potentialDecomposition}
\nabla_{(\!4)}A
=\nabla_{(\!4)}\!\cdot A+\nabla_{(\!4)}\!\wedge A .
\end{equation}
We define and denote these two parts of the field separately as
\begin{equation}\label{eq:SandF}
S\equiv\nabla_{(\!4)}\!\cdot A,
\qquad
G\equiv\nabla_{(\!4)}\!\wedge A= c^{-1}\vect E+I\vect B.
\end{equation}
Here, the electromagnetic field $G$ is represented by a spacetime
bivector, where $\vect E$ is the electric field strength,
$\vect B$ is the magnetic flux density, and $S$ defines the scalar
part of the field.

Thus, the field $F$ in Minkowski spacetime is expressed in terms of
its three-dimensional components ($S$, $\vect E$, $\vect B$) as
\begin{equation}\label{eq:PsiDef}
F\equiv\nabla_{(\!4)}A=S+G= S+c^{-1}\vect E+I\vect B.
\end{equation}

We now establish the relation between the potential and the
electromagnetic field in $\mathbb{R}^3$. For convenience, we first
evaluate the derivative in Eq.~\eqref{eq:PsiDef}:
\begin{align}\label{dA}
\gamma_0F\gamma_0=\gamma_0(\nabla_{(4)}A)\,\gamma_0=\Big(\frac{1}{c}\frac{\partial}{\partial t}+\vect e\,\nabla\Big)\nonumber\\
\Big(\frac1c\varphi+\vect A\Big)=\underbrace{\Big(\frac{1}{c^2}\frac{\partial}{\partial t}\varphi+\nabla\!\cdot \vect A\Big)}_{\text{scalar}}+\\ 
\underbrace{\frac1c\Big(\nabla\varphi+\frac{\partial}{\partial t}\vect A\Big)}_{\text{vector}}+\underbrace{I(\nabla\times \vect A)}_{\text{bivector}},\nonumber
\end{align}
where 
\begin{equation}\label{SF}
\gamma_0 F \gamma_0=S-c^{-1}\vect E+I\vect B \quad (\gamma_0\vect E\gamma_0=-\vect E),
\end{equation}
which follows from Eq.~\eqref{eq:PsiDef}.
Here, the relative basis 
\[e_k=\gamma_k\gamma_0, \quad e_k^2=1\] 
is
introduced with respect to the observer $\gamma_0$, and useful
geometric-algebra relations are employed (see
Appendix~\eqref{Ap}):
\begin{equation*}
\begin{gathered}
\nabla_{(4)} = \gamma_0\partial_0-\gamma_i\partial_i=(\partial_0-e_i\partial_i)\gamma_0=\gamma_0(\partial_0
+e_i\partial_i),\\
\Box = \nabla_{(4)}\nabla_{(4)}, \;
\gamma_\mu=\eta_{\mu\nu}\gamma_\nu,\; \gamma^i=-\gamma_i, \; \gamma^0\gamma_0=1. 
\end{gathered}
\end{equation*}

By matching Eqs.~\eqref{dA} and \eqref{SF} grade by grade, we obtain
the fields in $\mathbb{R}^3$:
\begin{equation}\label{S}
\begin{cases}
S=\dfrac{1}{c^2}\dfrac{\partial}{\partial t}\varphi+\nabla\!\cdot \vect A, \\
\vect E=-\nabla\varphi-\dfrac{\partial}{\partial t}\vect A,\\
\vect B=\nabla\!\times \vect A 
\end{cases}
\end{equation}
in terms of the electrodynamic potentials $\varphi$ and $\vect A$.

\subsection{System of Differential Equations of Electrodynamics}

Writing the wave equation \eqref{DA} as a decomposition by grade,
\begin{equation}\label{eq:Av}
\begin{gathered}
\nabla_{(4)}\nabla_{(4)}A\,\gamma_0=
\Big(\frac1c\frac{\partial}{\partial t}-\vect e\,\nabla\Big)\Big(\frac1c\frac{\partial}{\partial t}+\vect e\,\nabla\Big)\\
\Big(\frac1c\varphi+\vect A\Big)=\Big(\frac1c\frac{\partial}{\partial t}-\vect e\,\nabla\Big)
\big(S-\frac1c\vect E+I\vect B\big)=\\
\underbrace{\frac{1}{c}\Big(\frac{\partial}{\partial t}S+\nabla\cdot \vect E\Big)}_{\text{scalar}}+\underbrace{\Big(\nabla\times\vect B-\frac1{c^2}\frac{\partial}{\partial t}\vect E-\nabla S
\Big)}_{\text{vector}}+\\
\underbrace{\frac1{c}I\Big(\frac{\partial}{\partial t}\vect B+\nabla\times \vect E\Big)}_{\text{bivector}}-
\underbrace{ I\,\nabla\cdot \vect B}_{\text{pseudoscalar}}=\mu_0(c\rho+\vect j),
\end{gathered}
\end{equation}
and taking into account that $\gamma_0^2=1$, we separately equate the
scalar, vector, bivector, and pseudoscalar parts and obtain the
Maxwell system
\begin{subnumcases}{\label{eq:maxwell}}
\nabla\times\vect B=\mu_0 \vect j+\frac1{c^2}\frac{\partial}{\partial t}\vect E+\nabla S, \label{8a}\\
\nabla\times \vect E=-\frac{\partial}{\partial t}\vect B, \label{8b}\\
\nabla\cdot \vect B=0, \label{8c}\\
\nabla\cdot \vect E=\frac{1}{\varepsilon_0}\rho-\frac{\partial}{\partial t}S\label{8d},
\end{subnumcases}
augmented by the scalar field $S$.

The generalized Amp\`ere--Maxwell equation can be written
in terms of three current-density contributions. Introducing
the Maxwell displacement current density
\begin{equation} \label{eq:21}
 \vect j_d\equiv
 \varepsilon_0\frac{\partial\vect E}{\partial t}
\end{equation}
and the effective scalar field quantities
\begin{equation} \label{eq:22}
 \vect j_S\equiv\frac{1}{\mu_0}\nabla S,
 \qquad
 \rho_S\equiv-\varepsilon_0\frac{\partial S}{\partial t},
\end{equation}
we obtain
\begin{align}
 \nabla\times\vect B
 &=\mu_0\left(\vect j+\vect j_d+\vect j_S\right),\\
 \nabla\cdot\vect E
 &=\frac{1}{\varepsilon_0}\left(\rho+\rho_S\right).
\end{align}
By formal analogy with Maxwell's displacement current,
we refer to $\vect j_S$ as the scalar displacement current
density and to $\rho_S$ as the scalar displacement charge
density. These quantities are field contributions and do not
represent an additional conduction current or a new form of
electric charge.

\subsection{Wave Equations for the Electromagnetic Field}

The following wave equations, which provide a consistency check,
follow from the system \eqref{eq:maxwell}:
\begin{align}\label{eq:18}
\Box\vect E&=-\frac1\varepsilon_0\nabla \rho-\mu_0\frac{\partial}{\partial t}\vect j,\\
\Box\vect B&=\mu_0\nabla\!\times\vect j,\label{eq:19}\\
\Box S&=\mu_0\Big(\nabla\cdot\vect j+\frac{\partial}{\partial t}\rho\Big).\label{eq:20}
\end{align}

Indeed, applying the curl operator to Eq.~\eqref{8b}, using
Eqs.~\eqref{8a} and \eqref{8d}, and employing the identity
\[
\nabla\times(\nabla\times\vect E)=\nabla(\nabla\cdot\vect E)-\Delta\vect E,
\]
yields the wave equation for the electric field, Eq.~\eqref{eq:18}.
Similarly, Eq.~\eqref{eq:19} follows from Eq.~\eqref{8a} by using
Eqs.~\eqref{8b} and \eqref{8c}. Finally, applying the divergence
operator to Eq.~\eqref{8a} and using Eq.~\eqref{8d} gives
Eq.~\eqref{eq:20}.

It should be noted that the right-hand side of the last equation
contains no field quantities. Consequently, the scalar field $S$ is
excited directly by the charge and current sources rather than by
the electromagnetic field itself.

\section{Quadratic Operator and Energy--Momentum Tensor}\label{sec:energy_momentum}

Having constructed the system of field equations, we now proceed naturally to the study of the quadratic characteristics of the electromagnetic field, corresponding to the final stage of the logical sequence
\[
J\overset{*}{\longrightarrow}
A\overset{\partial}{\longrightarrow}
F
\overset{\mathcal T}{\longrightarrow}
\frac1{2\mu_0}F u\widetilde{F}: \{ w,\;\vect P,\;\vect g,\;\vect \sigma, \;\vect f \}.
\]

Unlike the field itself, energy and momentum remain unchanged under the transformation
\[
F\rightarrow -F,
\]
and therefore the corresponding physical quantity cannot depend linearly on the field.

Consequently, the simplest covariant object satisfying this requirement is a quadratic form constructed from the multivector
\begin{equation}\label{Fem}
F=S+G  \quad  \big(G =\vect E/c+I\vect B\big).
\end{equation}

In spacetime algebra, a second-rank tensor is naturally represented as a linear mapping of an arbitrary four-vector $a$ to a four-vector. We therefore seek the required quadratic form as a mapping
\[
a\longmapsto\mathcal T(a),
\]
which is linear in its argument $a$ and quadratic in the field $F$.

Below, we examine the algebraic structure of this mapping and show that its temporal and spatial components determine the energy density, energy flux, momentum density, and electromagnetic stress tensor.

\subsection{Definition and Properties of the Operator $\mathcal{T}$}

The construction of the operator must be based exclusively on the geometric structure of the multivector $F$. Since the product of two multivectors is already a quadratic function of the field, it is natural to consider a bilinear construction containing the multivector $F$ on the left and its reverse on the right. This construction ensures covariance while preserving linear dependence on the argument $a$.

The quadratic construction $\mathcal T(a)$ can naturally be expected to determine the energy characteristics of the field, since the multivector $F$ contains sufficiently complete information about the electromagnetic field.

Since the multivector $F$ contains scalar and bivector
components, the construction of the quadratic form must
account for their combined contribution. At the same time, the resulting object must preserve covariance and have a vector value for any choice of the argument $a$.

These requirements are satisfied by the bilinear construction
\begin{equation}\label{eq:Tmap}
\mathcal T(a)=\frac1{2\mu_0}F a\widetilde{F},
\qquad a\in\grade{1},
\end{equation}
where the symbol $\widetilde{F}$ denotes the reversion of the multivector.

Reversion is an antiautomorphism of geometric algebra:
\begin{equation*}
\widetilde{AB}=\widetilde B\,\widetilde A.
\end{equation*}
Consequently,
\begin{equation}
\widetilde{F}=S-G,
\label{eq:full-field-reversion}
\end{equation}
since the scalar and bivector satisfy $\widetilde S=S$ and $\widetilde G=-G$, respectively, while $S$ commutes with all elements of the algebra.

The operator $\mathcal T$ is linear in its argument:
\begin{equation}
\mathcal T(\alpha a+\beta b)
=
\alpha\mathcal T(a)+\beta\mathcal T(b),
\label{eq:quadratic-operator-linearity}
\end{equation}
but is quadratic in the field $F$.

The operator $\mathcal T(a)$ has the adjoint operator
\begin{align}\label{+operator}
\mathcal T^\dagger(a)
=\frac{1}{2\mu_0}
\widetilde{F}\,a\, F=\frac{1}{2\mu_0}
\left(S^2a-GaG\right) \nonumber\\
-\frac{S}{\mu_0}(G\cdot a),
\end{align}
where it is useful to employ the inner product defined in
Eq.~\eqref{scalarpro}.

The symmetric and antisymmetric parts of the operator are then
defined as
\begin{align}
\mathcal T(a)&= \frac{1}{2\mu_0}
\left(S^2a-GaG\right)+\frac{S}{\mu_0}(G\cdot a), \\
\mathcal T_{\mathrm s}(a)&=
\frac{1}{2\mu_0}\big(S^2a-GaG\big),\label{soperator} \\
\mathcal T_{\mathrm a}(a)&=\frac{S}{\mu_0}G\cdot a=\mathcal T(a)-\mathcal T_{\mathrm s}(a).\label{aoperator}
\end{align}
Here,
\begin{align*}
\mathcal T_{\mathrm s}(a)=\frac12\big(\mathcal T(a)+\mathcal T^\dagger(a)\big), \\ \mathcal T_{\mathrm a}(a)=\frac12\big(\mathcal T(a)-\mathcal T^\dagger(a)\big).
\end{align*}

The quadratic operator
\begin{equation}
\mathcal T(a)=\mathcal T_{\mathrm{s}}(a)+\mathcal T_{\mathrm{int}}(a) \quad (T_{\mathrm{int}}\equiv T_{\mathrm{a}}), \label{eq:operator-three-parts}
\end{equation}
can be structurally decomposed into independent contributions: the isotropic action of the scalar field, the standard electromagnetic-field operator, and a mixed scalar--bivector contribution. Using the expressions for $F$ in Eq.~\eqref{Fem} and $\widetilde{F}$ in Eq.~\eqref{eq:full-field-reversion}, we identify
\begin{equation}
\mathcal T_{\mathrm{em}}(a)=-\frac{1}{2\mu_0}GaG
\label{eq:electromagnetic-operator}
\end{equation}
as the quadratic operator of the electromagnetic bivector,
\begin{equation}
\frac{S^2}{2\mu_0}a
\label{eq:scalar-field-operator}
\end{equation}
as the intrinsic contribution of the scalar field, and
\begin{equation}
\mathcal T_{\mathrm{int}}(a)=\frac{S}{2\mu_0}(Ga-aG)\equiv \frac{S}{\mu_0}(G\cdot a)
\label{eq:interaction-operator}
\end{equation}
as the interaction between the scalar field and the electromagnetic bivector.

In the absence of the scalar field, $S=0$, Eq.~\eqref{eq:operator-three-parts} reduces to the standard electromagnetic form
\begin{equation}
\mathcal T(a)\big|_{S=0}=\mathcal T_{\mathrm{em}}(a).
\label{eq:maxwell-operator-limit}
\end{equation}

\subsection{Action of the Operator on an Observer}

Let $u$ be a unit timelike vector ($u^2=1$) characterizing an arbitrary inertial observer.

Any spacetime vector admits a unique decomposition into components parallel and orthogonal to $u$:
\begin{equation}
\mathcal T(u)=\big(\mathcal{T}(u)\cdot u\big)\,u+\big(\mathcal{T}(u)\wedge u\big)\,u
=\mathcal E(u)\,u+\mathcal P(u),
\label{eq:observer-decomposition}
\end{equation}
where
\begin{equation}
\mathcal E(u)=u\cdot\mathcal T(u)
\label{eq:observer-energy-density}
\end{equation}
is the scalar projection of $\mathcal T(u)$ onto the observer's time direction, while
\begin{equation}
\mathcal P(u)=\mathcal{T}(u)-\mathcal E(u)\,u=\big(\mathcal{T}(u)\wedge u\big)\,u
\label{eq:observer-spatial-part}
\end{equation}
is the part of the vector $\mathcal T(u)$ that is spatial relative to $u$.

The orthogonality of $u$ and $\mathcal P(u)$ follows directly from
Eq.~\eqref{eq:observer-decomposition}:
\begin{align}
u\cdot\mathcal P(u)\equiv0.
\label{eq:spatial-part-orthogonality}
\end{align}

The projection of $\mathcal T(u)$ onto $u$, namely $\mathcal E(u)$,
is interpreted as the field energy density measured by the observer,
whereas the orthogonal component $\mathcal P(u)$ is the spatial part
of the corresponding energy--momentum four-vector. Its specific
relation to the energy flux and momentum density is established
after the observer's reference frame has been chosen.

A coordinate representation of the operator arises only after an
orthonormal basis ${\gamma_\mu}$ has been selected. In this case, the
components of the linear mapping are defined by
\[
T_{\mu\nu}=\gamma_\mu\cdot \mathcal T(\gamma_\nu).
\]

Thus, the energy--momentum tensor arises as the coordinate
representation of the quadratic operator. The vector
$\mathcal{T}(\gamma_0)$ determines the energy density and energy flux
relative to the chosen observer, whereas the vectors
$\mathcal T(\gamma_i)$ $(i=1,2,3)$ determine the fluxes of spatial
momentum and the components of the stress tensor.

This approach provides a consistent transition from the
observer-independent geometric object, the field $F$, to the
quadratic operator and then to its coordinate representation,
without introducing additional assumptions. Once the components
$T_{\mu\nu}$ have been evaluated, the energy and momentum
conservation laws follow directly from the divergence of the
energy--momentum tensor.

\subsection{Temporal Projection}

We evaluate $\mathcal T_s(\gamma_0)$ in Eq.~\eqref{soperator}.

Using the commutation relations, we first calculate
\begin{align*}
G\,\gamma_0G={}&G (\gamma_0G\,\gamma_0)\,\gamma_0=\\
&{}\big(\vect E/c+I\vect B\big)\big(I\vect B-\vect E/c\big)\gamma_0=\\
&-\big(E^2/c^2+B^2+2I\vect B\wedge\vect E/c\big)\gamma_0=\\
&-\big(E^2/c^2+B^2+2\vect E\times\vect B/c\big)\gamma_0,
\end{align*}
and obtain
\begin{align}
\mathcal T_s(\gamma_0)=\frac{1}{2\mu_0}\big(S^2+E^2/c^2+B^2+2\vect E\times\vect B/c\big)\gamma_0.
\end{align}

Equation~\eqref{eq:interaction-operator} gives
\begin{align}
\mathcal T_{\mathrm{int}}(\gamma_0)=\frac{1}{\mu_0c}S\vect E\,\gamma_0,
\end{align}
where we have used
\[G\cdot\gamma_0\equiv(G-\gamma_0G\gamma_0)\,\gamma_0/2=\vect E\gamma_0/c\]
from Eq.~\eqref{scalarpro}, together with Eqs.~\eqref{eq:PsiDef}
and \eqref{SF}.

Thus,
\begin{align} \label{Tgamma}
\mathcal T(\gamma_0)=\mathcal T_{\mathrm{s}}(\gamma_0)+\mathcal T_{\mathrm{int}}(\gamma_0)=\big(w+\vect P/c)\gamma_0,
\end{align}
where
\begin{align}
w&=\frac12\left(\varepsilon_0E^2+\frac{B^2}{\mu_0}+\frac{S^2}{\mu_0}\right),\label{eq:u}\\
\vect{P}&=\frac1{\mu_0}\left(\vect E\!\times\vect B+S\vect E\right).\label{eq:P}
\end{align}
Equation~\eqref{Tgamma} is therefore derived from the
observer-independent field $F$, rather than introduced by analogy
with the conventional Poynting vector.

\subsection{All Spatial Projections}

To evaluate $\mathcal T_s(\gamma_i)$, we first transform the following vector expression:
\begin{equation}
\begin{split}
G\gamma_iG=(\vect E/c+I\vect B)\gamma_i(\vect E/c+I\vect B)=(\vect E/c+I\vect B)\\ 
\vect e_i\big(\gamma_0(\vect E/c+I\vect B)\gamma_0\big)\,\gamma_0=
(\vect E/c+I\vect B)\,\vect e_i\\
(-\vect E/c+I\vect B)\gamma_0=
\big(-\vect E\,\vect e_i \vect E/c^2-\vect B\,\vect e_i\\ \vect B+(\vect E\,\vect e_i\vect B-\vect B\,\vect e_i\vect E)I/c\big)\,\gamma_0
\end{split}
\end{equation}
and use the identities (see Eqs.~\eqref{EeE} and \eqref{EeB})
\begin{align*}
&\vect E\,\vect e_i\,\vect E\equiv
2(\vect E\cdot\vect e_i)\vect E-E^2\vect e_i,\\
&\vect E\,\vect e_i\,\vect B-\vect B\,\vect e_i\,\vect E\equiv -2I\vect e_i\cdot(\vect E\times \vect B).
\end{align*}

Using Eq.~\eqref{soperator}, we then obtain
\begin{align}
\mathcal T_s(\gamma_i){}={}&\bigg[\frac12\Big(\frac1\mu_0S^2-\varepsilon_0E^2-\frac1\mu_0B^2\Big)\vect e_i+\nonumber\\
&\varepsilon_0E_i\vect E+\frac1\mu_0B_i\vect B-\frac1{\mu_0c}\big(\vect E\times\vect B\big)_i\bigg] \gamma_0.
\end{align}
Expanding the quadratic operator in
Eq.~\eqref{eq:operator-three-parts} gives
\begin{align}\label{eq:Ti}
\mathcal T(\gamma_i)=\Big[\frac1{c\mu_0} \big(SE_i-(\vect E\!\times\vect B)_i\big) \notag+\nonumber\\
\big(\varepsilon_0E_i\vect E+\mu_0^{-1}B_i\vect B\big)+\mu_0^{-1}\Big(\vect e_i(S^2-\\
c^{-2}E^2-B^2)/2+S(\vect e_i\times\vect B)\Big)\Big]\gamma_0, \nonumber
\end{align}
where Eq.~\eqref{Scprod} has been used:
\begin{equation}
\mathcal T_{\mathrm{int}}(\gamma_i)=\frac{S}{\mu_0}{\Big(E_i/c+\vect e_i\times\vect B\Big)}\gamma_0.
\end{equation}

Thus, the single expression \eqref{eq:Ti} directly determines
$\mathcal T(\gamma_1)$, $\mathcal T(\gamma_2)$, and
$\mathcal T(\gamma_3)$.

\subsection{Definition of the Energy--Momentum Tensor}

Since the quadratic operator is vector-valued, we first define the
tensor by
\begin{equation}\label{tensor}
\mathcal T(\gamma_\nu)=T^\mu{}_\nu\,\gamma_\mu \quad \big(\mathcal T(\gamma_j)=T^0{}_j\,\gamma_0+T^i{}_j\,\gamma_i\big).
\end{equation}
Raising the lower tensor index gives
\begin{equation}\label{tensora}
T^{\mu\nu} =T^\mu{}_\alpha\,\eta^{\alpha\nu} \quad \big(T^{\mu0}=T^\mu{}_0, \quad T^{\mu i}=-T^\mu{}_i\big).
\end{equation}

For the spatial argument $\gamma_i$ in Eq.~\eqref{eq:Ti}, where the
Latin index $i$ denotes the input, the result may contain both a
temporal component $(T^0{}_i\,\gamma_0)$ and a spatial component
$(T^j{}_i\,\gamma_j)$. For example, the mixed component
\begin{align}
T^0{}_i=\frac{1}{\mu_0c}\big(SE_i-(\vect E\times\vect B)_i\big)   
\end{align}
is directed along $\gamma_0$.

We denote the spatial part of the tensor by
\begin{equation}
\sigma_{ij}\equiv T^i{}_j =-T^{ij},
\end{equation}
which can conveniently be written as
\begin{gather}\label{sigma}
\sigma_{ij}=(\sigma_{\mathrm M})_{ij} +
\frac{\delta_{ij}}{2\mu_0}S^2+\frac{S}{\mu_0}\epsilon_{ijk}B_k \\ \big((\sigma_{\mathrm M})_{ij}=(\sigma_{\mathrm M})_{ji}\big),\nonumber
\end{gather}
where
\begin{align}
(\sigma_{\mathrm M})_{ij}=\varepsilon_0
\Big(E_iE_j-\frac{E^2}2\delta_{ij}\Big)+
\frac1{\mu_0}\Big(B_iB_j \nonumber\\
-\frac{B^2}2\delta_{ij}\Big)
\end{align}
is the spatial Maxwell stress tensor.

The complete tensor can be shown to decompose structurally into the
standard Maxwell energy--momentum tensor, a purely scalar part, and
an interference part:
\begin{equation}
T^{\mu\nu}=T_{\mathrm M}^{\mu\nu}+T_{S}^{\mu\nu}+T_{\mathrm{int}}^{\mu\nu},
\end{equation}
given explicitly by
\begin{align}
T_{\mathrm M}^{\mu\nu}
&=\begin{pmatrix}
\left(\varepsilon_0E^2+\mu_0^{-1}B^2\right)/2
&c\varepsilon_0(\vect E\times\vect B)^{\mathsf T}
\\
c\varepsilon_0(\vect E\times\vect B)
&-\vect{\sigma}_{\mathrm M}
\end{pmatrix},\\
T_S^{\mu\nu}&=\frac{S^2}{2\mu_0}
\operatorname{diag}(1,-1,-1,-1),\\
T_{\mathrm{int}}^{\mu\nu}
&=\frac{S}{\mu_0}
\begin{pmatrix}
0&-E_1/c&-E_2/c&-E_3/c\\
E_1/c&0&-B_3&B_2\\
E_2/c&B_3&0&-B_1\\
E_3/c&-B_2&B_1&0
\end{pmatrix}.
\end{align}

\section{Conservation Laws}
\label{sec:conservation_laws}

Although the energy--momentum tensor follows from the quadratic
operator
\begin{equation*}
 \mathcal T(a)=\frac1{2\mu_0}Fa\widetilde F,  
\end{equation*}
the conservation laws are conveniently derived by taking the
divergence of the tensor $T^{\mu\nu}$. Energy and momentum then emerge
as the temporal and spatial components of a single four-dimensional
equation.

For convenience, we choose the following block-matrix representation
of the energy--momentum tensor:
\begin{equation}
T^{\mu\nu}=\begin{pmatrix}w&c\,\vect g^{\mathsf T}\\
\vect P/c&-\vect \sigma
\end{pmatrix},
\end{equation}
where $w$ is the total field energy density, $\vect P$ is the energy
flux, and $\vect g$ is the momentum density:
\begin{align}
w&=\frac12\left(\varepsilon_0E^2+\frac{B^2}{\mu_0}+\frac{S^2}{\mu_0}
\right),\\
\vect P&=\frac1\mu_0\vect E\times\vect B
+\frac1\mu_0S\vect E,\\
\vect g&=\varepsilon_0\left(
\vect E\times\vect B-S\vect E
\right)\label{impulsdensity}.
\end{align}

\subsection{Unified Balance Equation}

We write the four-force density $f^\nu$ acting on charges and currents
as
\begin{equation*}
\partial_\mu T^{\mu\nu}=-f^\nu.
\end{equation*}
The minus sign represents the loss of field energy and momentum.

This component form is considerably more convenient than the
operator representation
\begin{equation*}
\partial\cdot\mathcal T=-\vect f.
\end{equation*}

\subsection{Energy Balance}

For $\nu=0$,
\begin{equation} \label{diver}
\partial_\mu T^{\mu0}=-f^0.
\end{equation}

Taking into account
\begin{equation}
T^{00}=w,\qquad
T^{i0}=P_i/c,
\end{equation}
together with the field equations \eqref{eq:maxwell}, we evaluate the
divergence $(\partial w/\partial t+\nabla\!\cdot\!\vect P)/c$ and
obtain the source term, or the right-hand side of Eq.~\eqref{diver},
as
\begin{align}
f^0=\frac1c\vect j\cdot\vect E-c\rho S,
\end{align}
where $f^0$ has the dimensions of force density.

The energy conservation law therefore takes the form
\begin{align}\label{eq:energy}
\frac{\partial w}{\partial t}+\nabla\cdot\vect P=-\vect j\cdot\vect E+c^2\rho S,
\end{align}
where the right-hand side determines the volumetric rate of energy
exchange between the field and its sources.

The conservation law \eqref{eq:energy} can also be derived directly
from the field equations \eqref{eq:maxwell} by using the identity
\begin{equation*}
\nabla\cdot(\vect E\times \vect B)
\equiv \vect B\cdot(\nabla\!\times \vect E)-
\vect E\cdot(\nabla\!\times \vect B),
\end{equation*}
together with Eqs.~\eqref{8a}, \eqref{8b}, and \eqref{8d}. This
provides an independent verification of the temporal projection of
the quadratic operator.

\subsection{Momentum Balance and Force}

For $\nu=i$, we have
\begin{align}\label{impul}
\partial_\mu T^{\mu i}=\partial_0T^{0i}+\partial_jT^{ji}=-f^i,
\end{align}
together with
\begin{equation}
T^{0i}=c\,g_i.
\end{equation}

The spatial part is related to the stress tensor by
\begin{equation*}
T^{ji}=-\sigma_{ji}.
\end{equation*}

Equation~\eqref{impul} then gives the following expression for the
force:
\begin{equation*}
\frac{\partial g_i}{\partial t}-\partial_j\sigma_{ji}=-f_i,
\end{equation*}
or, in three-dimensional form,
\begin{equation}\label{eq:momentumgeneric}
 \vect f=\nabla\!\cdot\vect \sigma-\frac{\partial}{\partial t}\vect g
\end{equation}
with $\big(\partial_j\sigma_{ji}
\equiv(\nabla\cdot\vect\sigma)_i\big)$.

Evaluating the right-hand side using the expressions for
$\vect\sigma$ and $\vect g$ yields the generalized Lorentz force
density
\begin{equation}\label{eq:forcecandidate}
\vect f=\rho\vect E+\vect j\!\times\vect B-S\vect j.
\end{equation}

Indeed, Eq.~\eqref{sigma}, together with the property of the
Levi-Civita symbol $\epsilon_{jik}=-\epsilon_{ijk}$, gives
\begin{equation}
\nabla\cdot\vect\sigma=\nabla\cdot\vect\sigma_{\mathrm M} +
\frac{1}{2\mu_0}\nabla S^2-\frac{1}{\mu_0}\nabla\times(S\vect B),
\end{equation}
since the divergence of the linear antisymmetric contribution to the
tensor is
\[
\begin{aligned}
(\nabla\cdot\vect\sigma)_i^{SB}=\frac{1}{\mu_0}\partial_j
\left(S\epsilon_{jik}B_k\right)=-\frac{1}{\mu_0}\partial_j
\left(S\epsilon_{ijk}B_k\right)=\\
-\frac{1}{\mu_0}\left[\nabla S\times\mathbf B+S\nabla\times\mathbf B\right]_i=-\frac{1}{\mu_0}[\nabla\times(S\vect B)]_{i}.
 \end{aligned}
 \]

On the other hand, direct evaluation of the time derivative using
the field equations \eqref{eq:maxwell} gives
\begin{equation}
\begin{aligned}
\frac{\partial}{\partial t}\vect g=\nabla\bigg( \varepsilon_0
\Big(\vect E\vect E-\frac12E^2\Big)+\frac1{\mu_0}
\Big(\vect B\vect B-\\
\frac12B^2\Big)\bigg)-\frac{1}\mu_0\Big(\nabla S\times\vect B+S\nabla\times\vect B-\\
\frac12\nabla S^2\Big)-\vect f. 
\end{aligned}
\end{equation}

Thus, taking into account the spatial block \eqref{impul}, the energy
and momentum balance equations can be combined into the unified form
\begin{equation}\label{balans}
\partial_\mu T^{\mu\nu}=-f^\nu.
\end{equation}

For $S=0$, the antisymmetric part vanishes, leaving the standard
symmetric Maxwell tensor, while Eq.~\eqref{eq:forcecandidate}
reduces to the familiar Lorentz force density
\begin{equation}\label{eq:Lorents}
\vect f_L=\rho\vect E+\vect j\!\times\vect B.
\end{equation}

\section{Structure of a Free Scalar Wave}
\label{sec:s_wave}

A free scalar field satisfies the homogeneous wave equation
\begin{equation}
 \Box S=0.
 \label{eq:free_s_wave}
\end{equation}
Characteristic solutions of the homogeneous equation can be
represented in spherical and plane-wave forms. These representations
describe different aspects of the same process and should not be
regarded as mutually exclusive.

For a localized source, the natural solution is the retarded
spherically symmetric form
\begin{equation}
 S(\vect r,t)
 =
 \frac{1}{r}\,
 s\left(t-\frac{r}{c}\right),
 \qquad r>0,
 \label{eq:spherical_s_wave}
\end{equation}
where the function $s$ is determined by the time dependence of the
source. This form is directly related to the retarded Green function
of the wave operator. In three spatial dimensions, its support lies
on the light cone, expressing Huygens' principle: a disturbance
propagates away from the source at the speed $c$.

Solution~\eqref{eq:spherical_s_wave} is valid outside the source
region. Its singularity at $r=0$ shows that a strictly spherical
outgoing wave cannot exist without a source or an appropriate
boundary condition. At large distances, within a region whose
dimensions are small compared with the radius of the wavefront, the
spherical wave is locally approximated by a plane wave.

The plane-wave representation used below is therefore not an
assertion of the existence of an infinite uniform wave, but a local
model of a freely propagating wavefront. It provides the most
transparent means of determining the energy density, energy flux,
and characteristic properties of the longitudinal mode.

Let the source be absent:
\begin{equation}
\rho=0,\qquad \vect j=0.    
\end{equation}

The field then satisfies the extended system
\begin{align}
&\nabla\!\cdot\vect E=-\frac{\partial S}{\partial t},\\
&\nabla\times\vect B-\frac1{c^2}
\frac{\partial\vect E}{\partial t}
=\nabla S.
\end{align}

\subsection{Plane S-Wave}

For a wave propagating along the unit vector $\vect n$, let
\begin{equation}
S=S(\xi),
\qquad
\xi=t-\frac{\vect n\cdot\vect r}{c}.
\end{equation}

The longitudinal solution has the form
\begin{equation}
\vect E=cS\,\vect n,
\qquad
\nabla\times\vect B=0.
\end{equation}

For a plane-wave dependence $\vect B=\vect B(\xi)$,
the conditions $\nabla\times\vect B=0$ and
$\nabla\cdot\vect B=0$ imply $d\vect B/d\xi=0$.
Thus, $\vect B$ can contain only a constant background
component, which is set to zero for the free $S$-wave
considered here.

Indeed,
\begin{equation}
\nabla\!\cdot\vect E
= -\frac{1}{c}\frac{d}{d\xi}(cS)=
 -\frac{dS}{d\xi}=
-\frac{\partial S}{\partial t},
\end{equation}
while the Ampère equation gives $(\vect B=0)$
\begin{equation}
-\frac1{c^2}\frac{\partial\vect E}{\partial t}
= -\frac{\vect n}{c}\frac{dS}{d\xi}=
\nabla S.
\end{equation}

Thus, the electric field is directed along the propagation direction:
\begin{equation}
\vect E\parallel\vect n,
\end{equation}
while the rotational magnetic-field component is absent. This
qualitatively distinguishes the $S$-wave from a conventional
electromagnetic wave, for which:
\begin{equation}
\vect E\perp\vect n, \quad \vect B\perp\vect n, \quad \vect E\perp\vect B.
\end{equation}

For the solution obtained,
\begin{equation}
E^2=c^2S^2,
\qquad
B=0.
\end{equation}

The energy density is
\begin{equation}
w=\frac12\left(\varepsilon_0E^2+\frac{S^2}{\mu_0}\right)
=\frac{S^2}{\mu_0}.
\end{equation}

The energy flux is
\begin{equation}\label{PS}
\vect P
=\frac1{\mu_0}\vect E\times\vect B
+\frac1{\mu_0}S\vect E
=\frac{cS^2}{\mu_0}\vect n.
\end{equation}

Consequently,
\begin{equation}
\vect P=cw\,\vect n.
\end{equation}

The momentum density is
\begin{equation}
\vect g=\varepsilon_0
\left(\vect E\times\vect B-S\vect E\right)
= -\varepsilon_0cS^2\vect n=
-\frac{w}{c}\vect n.
\end{equation}

Thus,
\begin{equation}
\vect P=-c^2\vect g,
\qquad
|\vect P|=cw.
\end{equation}

The longitudinal $S$-wave therefore transports energy at
the speed $c$ and possesses a nonzero momentum density; it
is neither a Coulomb-field solution nor a near-field
solution.

\subsection{Spherical Wave}

For experimental purposes, a spherically symmetric solution is also
required:
\[
S(r,t)=\frac1r f\left(t-\frac rc\right).
\]

The corresponding radial electric field (for $\vect B=0$) has the
form
\[
\vect E(r,t)
=\left[
\frac{c}{r}f\left(t-\frac rc\right)
+\frac{c^2}{r^2}F\left(t-\frac rc\right)
\right]\vect e_r,
\]
$F'(\xi)=f(\xi)$.

Of particular importance is the decomposition
\[
\vect E=\vect E_{\mathrm{wave}}+\vect E_{\mathrm{near}},
\]
where
\[
\vect E_{\mathrm{wave}}=\frac{c}{r}f\left(t-\frac rc\right)\vect e_r=cS\,\vect e_r
\]
decreases as $r^{-1}$ and transports energy, whereas
\[
\vect E_{\mathrm{near}}=\frac{c^2}{r^2}F\left(t-\frac rc\right)\vect e_r
\]
is the near-field component.

In the far field,
\[
\vect E\simeq cS\,\vect e_r,
\]
and therefore the energy flux decreases as
\[
|\vect P|
\simeq
\frac{c}{\mu_0r^2}f^2\left(t-\frac rc\right),
\]
in accordance with Eq.~\eqref{PS}, while the total power passing
through a sphere remains finite:
\[
P_{\mathrm{tot}}
= \int d\vect a\cdot \vect P\simeq\frac{4\pi c}{\mu_0}
f^2\left(t-\frac rc\right).
\]

The spherical solution determines the far-field structure to be
sought when the $S$-component is excited by a localized effective
source.

Unlike the locally plane-wave approximation, the spherical solution
contains both radiative and near-field components:
\[
S=\frac{f(\xi)}r,
\qquad
\vect E=
\left(
\frac{cf(\xi)}r+\frac{c^2F(\xi)}{r^2}
\right)\vect e_r.
\]
The energy density of this solution separates into
\begin{itemize}
\item a radiative term of order ($r^{-2}$);
\item a mixed term of order ($r^{-3}$);
\item a near-field term of order ($r^{-4}$).
\end{itemize}

In the near-field region, the ratio
\[
|\vect P|/w,
\]
is generally less than $c$, because part of the energy is stored in
the bound field. This makes it possible to distinguish freely
propagating radiation from a bound longitudinal near-field
component.

\subsection{Complete Energy--Momentum Tensor of a Plane S-Wave}

For a pure traveling wave, we have ($w_M=w_S$)
\[
w=w_M+w_S=\frac{S^2}{\mu_0},
\quad
\vect P=cw\,\vect n,
\quad
\vect g=-\frac{w}{c}\vect n.
\]

It should be emphasized that, although the scalar component $S$
satisfies a separate wave equation, the propagating $S$-mode also
includes an associated longitudinal electric component. The field
equations relate it to the longitudinal electric field through
\begin{equation}
 \vect E_\parallel=cS\vect n,\qquad \vect B_S=0.
\end{equation}
Here, $\vect B_S=0$ does not imply the absence of a magnetic field in
a general configuration containing both the $S$-sector and the
Maxwell sector.

The energy--momentum tensor of this mode therefore contains scalar,
Maxwell, and interference parts. Their separation is algebraically
meaningful; however, the individual parts do not describe
independently propagating subsystems.

The spatial part must have the form
\begin{equation}
T^{\mu\nu}
=T_{\mathrm M}^{\mu\nu}
+T_S^{\mu\nu}
+T_{\mathrm{int}}^{\mu\nu}
=w\begin{pmatrix}
1 & -\vect n^{\mathsf T}\\
\vect n & -\vect n\vect n^{\mathsf T}
\end{pmatrix},
\label{eq:plane_s_full_tensor}
\end{equation}
where $T^{0i}=cg_i=-wn_i$, $T^{i0}=P_i/c=wn_i$,
$(T^{\mathrm{int}})_{ij}=0$, and $(T_S)^{ij}=-\delta_{ij}w/2$,
\begin{equation*}
(T_M)^{ij}=-(\sigma_M)_{ij}=-w n_in_j+\frac w2\delta_{ij}.
\end{equation*}

Consequently, choosing the propagation direction along the $z$ axis
($\vect n=\vect e_z$), the complete tensor can be written as
\begin{equation}
T^{\mu\nu}=w
\begin{pmatrix}
1&0&0&-1\\
0&0&0&0\\
0&0&0&0\\
1&0&0&-1
\end{pmatrix}.
\label{eq:plane_s_tensor_z}
\end{equation}

\subsection{Energy Transport Velocity}

The energy transport velocity is defined by the ratio
\begin{equation}
 \vect v_{\mathrm{en}}=\vect P/w=c\vect n.
\end{equation}
The wave equation $\Box S=0$ yields the dispersion relation
$\omega=c|\vect k|$; therefore, the phase and group velocities of the
free mode are also equal to $c$.

\subsection{Polarization Properties}

A scalar wave does not possess two independent transverse
polarization states. Its field is uniquely related to the propagation
direction:
\[
\vect E=cS\,\vect n.
\]

Therefore, a free plane $S$-wave
\begin{itemize}
\item does not exhibit conventional linear or circular polarization;
\item has no magnetic component.
\end{itemize}

\section{Source of Scalar S-Waves} \label{sourcesection}

The generation of scalar $S$-waves may be associated with
impressed forces and local charge redistribution between
interacting physical subsystems. For the effective four-current
of a selected open subsystem,
\begin{equation}\label{eq.KS}
 \mathcal K = \frac{\partial\rho}{\partial t}
 +\nabla\!\cdot\vect j = \partial_\mu J^\mu,
\end{equation}
In general, $\mathcal K$ need not vanish for an open subsystem.

This does not imply a violation of charge conservation.
If
\begin{equation} \label{eq:103} J_{\mathrm{full}}^\mu=J^\mu+J_{\mathrm{rest}}^\mu
\end{equation}
is the total physical current of the closed system, then
\begin{equation} \label{eq:104}
\partial_\mu J_{\mathrm{full}}^\mu=0,
\qquad
\partial_\mu J_{\mathrm{rest}}^\mu=-\mathcal K.
\end{equation}

According to the scalar wave equation \eqref{eq:20} derived in
Section \ref{sec:field_equations},
the quantity $\mathcal K$ acts as the local source of the scalar field.

Using the scalar displacement quantities introduced in
Eqs.~\eqref{eq:21} and \eqref{eq:22}, we obtain
\begin{equation}
 \frac{\partial\rho_S}{\partial t}
 +\nabla\cdot\vect j_S
 =-\frac{1}{\mu_0}\Box S=-\mathcal K.
\end{equation}
Consequently,
\begin{equation} \label{totcur}
 \frac{\partial}{\partial t}(\rho+\rho_S)
 +\nabla\cdot(\vect j+\vect j_S)=0.
\end{equation}
Equation \eqref{totcur} expresses the continuity of the effective
source--field combination. Independently, Eqs. \eqref{eq:103} and \eqref{eq:104} ensure conservation of the total physical charge of
the closed system.

In a region where $\mathcal K=0$, the scalar field satisfies the homogeneous wave equation
\begin{equation}
 \Box S=0.
\end{equation}

This condition excludes a local source; however, the homogeneous
equation does not imply that the $S$-field is trivial. The solution
$S=0$ is selected only by the corresponding initial and boundary
conditions.

An illustrative impulsive source model is provided by a charged
particle passing through a small aperture in an infinite conducting
screen. Two physical subsystems interact in this problem: the moving
charge and the conducting screen with the surface charges and
currents induced in it. If field penetration through the small
aperture is neglected, then, for an observer behind the screen, the
charge appears to be created at the instant of passage. Before that
instant, neither the charge itself nor its self-field was present in
the observation region; afterward, the particle appears immediately
with a prescribed velocity. The self-field of the charge, however,
cannot become established instantaneously throughout the entire
region and is restored causally at the speed of light.

Taking the instant at which the particle passes through the aperture
as the time origin, the effective charge and current densities in
the region behind the screen can be represented as
\begin{align}
 \rho_{+}(\vect r,t) = q\,\delta^{(3)}(\vect r-\vect v t)\theta(t),
 \;\! \vect j_{+}(\vect r,t) = \nonumber\\
 q\vect v\,\delta^{(3)}(\vect r-\vect v t)\theta(t).
\end{align}
If the particle passes through the coordinate origin at $t=0$, we
obtain
\begin{equation}
 \mathcal K_{+}
 = \frac{\partial\rho_{+}}{\partial t}
 +\nabla\!\cdot\vect j_{+}
 = q\,\delta^{(3)}(\vect r)\delta(t)\ne0,
\end{equation}
since the following equalities hold in the sense of distributions:
\begin{align*}
\frac{\partial\rho_{+}}{\partial t}
=-q\,\theta(t)\,\vect v\!\cdot\nabla
 \delta^{(3)}(\vect r-\vect vt)\\
+q\delta^{(3)}(\vect r-\vect vt)\delta(t),\\
\nabla\!\cdot\vect j_{+}=q\,\theta(t)\,\vect v\!\cdot\nabla
 \delta^{(3)}(\vect r-\vect vt).
\end{align*}

Thus, for the selected subsystem, the passage of the particle through
the physical boundary of the shielded region appears as a local
creation of charge. In the complete physical system, the appearance
of the charge behind the screen is accompanied by its disappearance
in front of the screen; therefore, the global conservation law for
electric charge is not violated.

In the proposed model, the localized quantity $\mathcal K_{+}$
enters the right-hand side of the equation
\begin{equation}
 \Box S=\mu_0\mathcal K_{+}
\end{equation}
and acts as the source of a pulsed $S$-field accompanying the causal
restoration of the charge's self-field.

\section{Discussion}
\label{sec:discussion}

The system of equations \eqref{eq:maxwell} preserves the
correspondence principle: for $S=0$, it reduces to Maxwell
electrodynamics, while the expressions for the energy density,
energy flux, momentum, and stresses reduce to their corresponding
standard forms. Therefore, introducing the scalar component does not
alter known solutions within their domains of applicability, but
extends the class of admissible field states.

The principal physical distinction of the extended system is the
possibility of a free scalar $S$-wave accompanied by a longitudinal
electric component. The resulting energy--momentum tensor shows that
such a wave is capable of transporting energy and momentum. The
field $S$ may therefore have potentially observable dynamical
manifestations and does not merely amount to an alternative
representation of the standard electromagnetic variables.

A detailed physical interpretation of the nonsymmetric
mixed sector of the energy--momentum tensor, including the
relation between energy flux and momentum density in
combined Maxwell and $S$-field configurations, requires a
separate study and is left for future work.

The equation
\begin{equation*}
 \Box S=\mu_0\mathcal K
\end{equation*}
relates the excitation of the scalar field to the nonzero divergence
of the effective impressed current of an open subsystem. The most
natural conditions for such a source arise in rapid charge
redistribution processes, including the switching on and off of
sources, pulsed charging of conductors, capacitor discharge, passage
of a charge through an aperture in a screen, and transient processes
in antennas. After the source ceases to act, the excited field can
propagate through free space as a solution of the homogeneous
equation $\Box S=0$.

The theoretical significance of the results lies in providing a
consistent description of the scalar degree of freedom of the
electromagnetic field, from its geometric origin and field equations
to its energy characteristics and excitation mechanism. Maxwell
electrodynamics is retained as a special case of the extended system,
while the possible existence of an $S$-wave leads to testable
physical consequences.

At the present stage, the practical significance of this work lies
primarily in identifying the conditions required for the generation
and detection of $S$-waves. The equation
$\Box S=\mu_0\mathcal K$ provides a criterion for selecting a source,
while the derived expressions for energy, energy flux, and momentum
make it possible to formulate requirements for the generator,
receiver, and measurement procedure. This provides a basis for
analyzing pulsed sources and transient processes in antennas and
conducting structures, as well as for designing experiments capable
of distinguishing a predicted $S$-signal from transverse radiation
and known longitudinal fields.

The possibility of deliberately exciting a longitudinal electric
component is of interest for antenna technology, signal and energy
transmission, shielding studies, and the interaction of radiation
with plasmas and dispersive media. The extended system may also be
useful in analyzing rapid transient processes in which the
steady-state field of a source has not yet been established
throughout space.

Because longitudinal fields are known to occur in the near-field
region of antennas, waveguides, plasmas, and transition radiation,
and can be described by standard electrodynamics, experimental
identification of an $S$-wave must be based on a combination of
signatures: persistence of the longitudinal component beyond the
near-field region, propagation of the disturbance at the speed of
light, transport of energy and momentum, and dependence of the
signal amplitude on the magnitude and spacetime structure of the
source $\mathcal K$.

Of particular importance is the design of an experiment capable of
separating the contribution of the $S$-field from transverse
electromagnetic radiation and the quasistatic fields of the source.
This may involve spatial separation of the field components,
measurement of the signal dependence on distance, receiver
orientation, and screen parameters, as well as comparison of the
result with the complete solution of the corresponding problem in
Maxwell electrodynamics.

The design of a specific generator and the development of
quantitative criteria for such an experiment remain subjects for
future investigation.

\section{Conclusion}
\label{sec:conclusion}

Using spacetime algebra, an extension of classical electrodynamics
has been constructed in which the unified field contains an
observer-independent scalar component $S$ in addition to the
electromagnetic bivector part. The resulting system preserves the
correspondence principle and reduces completely to Maxwell's
equations when $S=0$.

It has been shown that the free scalar field satisfies a wave
equation and is accompanied by a longitudinal electric component.
A generalized expression for the Lorentz force density has been
obtained, containing an additional contribution from the scalar
field:
\begin{equation}
 \vect f
 =
 \rho\vect E+\vect j\times\vect B-S\vect j.
\end{equation}
The additional term $-S\vect j$ defines a new mechanism of momentum
exchange between the extended field and the physical current.

The energy--momentum tensor of the extended field has been
constructed from the quadratic operator. The energy density, energy
flux, momentum density, and stress tensor, including the
contributions of the scalar field, have been derived from it. Local
energy and momentum conservation laws have also been obtained,
establishing the balance between changes in the field energy and
momentum, their transport through space, and their interaction with
physical sources. The $S$-wave is thereby shown to represent a
dynamical field state capable of transporting energy and momentum
and exchanging them with matter.

The excitation mechanism of the scalar field has been identified.
The source of an $S$-wave is the nonzero local divergence of the
effective impressed current of an open subsystem:
\begin{equation}
 \Box S=\mu_0\mathcal K,
 \qquad
 \mathcal K=
 \frac{\partial\rho}{\partial t}
 +\nabla\!\cdot\vect j.
\end{equation}
The conservation law for the total electric charge of the closed
system remains satisfied. Outside the source region, the field
propagates as a free solution of the homogeneous wave equation.

The results provide a theoretical basis for modeling the generation
and reception of $S$-waves and establish the observable signatures
required to distinguish them from transverse electromagnetic
radiation and known longitudinal fields. Experimental tests of these
predictions must determine whether the scalar component constitutes
an independent physical degree of freedom of the electromagnetic
field.
\bibliography{Ref} 
\begin{appendices}
	\numberwithin{equation}{section}

\section{Clifford Algebra and Operator Decompositions}
\label{Ap}

\subsection{Spacetime Basis}

For a covariant description of electrodynamics, we employ the
spacetime algebra $\mathrm{Cl}(1,3)$, which is the geometric algebra
of Minkowski spacetime. This formalism unifies vectors, bivectors,
spinors, and Lorentz transformations within a single algebraic
structure
\cite{hestenes1966spacetime,hestenes2003,Doran}.

We introduce four basis elements
$(\gamma_0,\gamma_1,\gamma_2,\gamma_3)$ as generators of the Clifford
spacetime algebra $\mathrm{Cl}(1,3)$. They satisfy
\[
\gamma_\mu \gamma_\nu + \gamma_\nu \gamma_\mu = 2 g_{\mu\nu},
\]
where $g_{\mu\nu}=\mathrm{diag}(+1,-1,-1,-1)$. Basis vectors with
different indices anticommute:
\begin{equation}\label{gam}
\gamma_\mu\gamma_\nu=
-\gamma_\nu\gamma_\mu\quad (\mu\neq\nu).
\end{equation}
The squares of the basis vectors define the Minkowski metric
$g_{\mu\nu}$:
\[
\gamma_0^2=+1,\;
\gamma_1^2=\gamma_2^2=\gamma_3^2=-1.
\]

Any four-vector in STA is represented as
$a = a^\mu \gamma_\mu$.
Its square automatically gives the Minkowski interval:
\[
a^2=(a^0)^2-(a^1)^2-(a^2)^2-(a^3)^2.
\]

In STA, the timelike direction $\gamma_0^2=+1$ differs from the
spacelike directions $\gamma_i^2=-1$.

This is not a property of a particular vector, but a property of the
spacetime geometry itself: a unit timelike vector in STA has positive
squared norm, whereas a unit spacelike vector has negative squared
norm.

\subsection{Geometric Product}

STA introduces the more general geometric product $ab$, which
decomposes into symmetric and antisymmetric parts:
\begin{align}
ab = a\cdot b + a\wedge b  \label{geomprod},
\end{align}
where
\begin{gather}
a \cdot b = \frac{1}{2}(ab + ba), \label{simprod}\\
a \wedge b = \frac{1}{2}(ab - ba) \label{asprod}.
\end{gather}
The first part of $ab$ is a scalar, or inner product, whereas the
second is the outer product, or a bivector representing an oriented
plane element spanned by the vectors $a$ and $b$.

\subsection{Grade Structure}

A multivector decomposes by grade as
\[
M = \langle M \rangle_0 + \langle M \rangle_1 + \langle M \rangle_2 + \langle M \rangle_3 + \langle M \rangle_4,
\]
where the corresponding components are a scalar, vector, bivector,
pseudovector, and pseudoscalar.

The following useful relations hold for a vector and a bivector:
\begin{gather} 
G\cdot u=\frac12(Gu-uG)=-u\cdot G, \label{scalarpro}\\
(a\wedge b)\cdot u=a(b\cdot u)-b(a\cdot u),\\
u\cdot(G\cdot u)=0.
\end{gather}

\subsection{Pseudoscalar and Duality}

A special role is played by the pseudoscalar
\[
I=\gamma_0\gamma_1\gamma_2\gamma_3,
\qquad I^2=-1,
\]
which defines the duality operation. Thus, the formalism does not
require the introduction of an independent imaginary unit.

Multiplication of a bivector by the pseudoscalar $I$ produces its
dual bivector:
\[
G\longmapsto IG.
\]

\subsection{Spatial Pauli Basis}
\label{paul}

By fixing an observer ($\gamma_0$), we can introduce the relative
spatial basis
\[
e_k=\gamma_k\gamma_0 \quad (k=1,2,3).
\]
This basis can be treated as an ordinary Euclidean three-dimensional
basis $(e_k^2=1)$:
\[
e_ie_j=\delta_{ij}+I\,\epsilon_{ijk}\,e_k, \ I=e_1e_2e_3=\gamma_0\gamma_1\gamma_2\gamma_3,
\]
where $\delta_{ij}$ is the Kronecker delta and $\epsilon_{ijk}$ is
the Levi-Civita symbol.

\subsection{d'Alembert Differential Operator}

We write the vector differential operator (Dirac operator) as
\begin{equation}\label{D}
\nabla_{(4)} =\gamma_0\partial_0 - \gamma_i \partial_i= (\partial_0 - \nabla)\gamma_0=\gamma_0(\partial_0 + \nabla),
\end{equation}
where the commutation property of $\gamma_0$ has been used, and
$\nabla$ denotes the three-dimensional nabla operator
\begin{equation}\label{N}
\nabla=e_i\partial^i=e_1\partial_x+e_2\partial_y+e_3\partial_z.
\end{equation}

The product of two operators $\nabla_{(4)}$ in
$\mathrm{Cl}(1,3)$ yields the scalar d'Alembert operator
\begin{equation*}\label{Dal}
\Box= \Big(\frac{1}{c}\frac{\partial}{\partial t} - \nabla\Big)\!\gamma_0\gamma_0\!\Big(\frac{1}{c}\frac{\partial}{\partial t} + \nabla\Big)=\Big(\frac{1}{c^2}\frac{\partial^2}{\partial t^2} - \nabla^2\Big),
\end{equation*}
corresponding to Minkowski spacetime, by virtue of the commutation
relation in Eq.~\eqref{D} and the equality $\gamma_0^2=1$.

\subsection{Key Operator Decompositions}
\label{key}

For spatial vector fields, we prove the decomposition
\begin{equation}\label{pA}
\nabla \vect A\equiv\nabla\cdot \vect A+I(\nabla\times\vect A),
\end{equation}
starting from the definition of the geometric product
\begin{equation*}
\nabla \vect A=(e_i \partial_i)(A_j e_j)=
(\partial_i A_j)e_i e_j
\end{equation*}
and the basis-element relation
\[
e_i e_j=e_i\cdot e_j+e_i\wedge e_j.
\]
It follows that
\begin{equation}\label{NA}
\nabla \vect A=
(\partial_iA_j)(e_i\cdot e_j)
+
(\partial_iA_j)(e_i\wedge e_j).
\end{equation}
The first term can be written as the inner product
\begin{equation}\label{NA1}
(\partial_iA_j)(e_i\cdot e_j)=
(\partial_iA_j)\delta_{ij}=
\partial_iA_i=
\nabla\cdot \vect A,
\end{equation}
whereas the second term is expressed through the outer product:
\[
(\partial_iA_j)(e_i\wedge e_j)=
\nabla\wedge \vect A,
\]
which, in three-dimensional algebra, is related to the curl by
\begin{equation}\label{NA2}
\nabla\wedge \vect A=
I(\nabla\times \vect A).
\end{equation}

Substituting the expressions obtained in Eqs.~\eqref{NA1} and
\eqref{NA2} into Eq.~\eqref{NA} yields the identity \eqref{pA}.

This identity reflects the decomposition of the geometric product
into scalar and bivector parts.

We next consider a vector decomposition using the geometric-product
decomposition of $\vect e\vect E$ in Eq.~\eqref{simprod}:
\begin{equation} \label{EeE}
\vect E\,\vect e \vect E=\vect E\big(-\vect E\vect e+2(\vect e\cdot\vect E)\big)=2\vect E(\vect e\cdot \vect E)-E^2\vect e.
\end{equation} 

Using the geometric product \eqref{geomprod}, we first obtain
\begin{align*}
\vect E\,\vect e\vect B=\vect E\,(\vect e\cdot \vect B)+I\vect E(\vect e\times\vect B)=\vect E\,(\vect e\cdot\vect B)+ \\
{}I\vect E\cdot(\vect e\times \vect B)-\vect E\times(\vect e\times \vect B)=\vect E\,(\vect e\cdot \vect B)+\\
I\vect E\cdot(\vect e\times\vect B)-\vect e\,(\vect E\cdot \vect B)+\vect B\,(\vect E\cdot \vect e). 
\end{align*}
Interchanging $\vect E\longleftrightarrow\vect B$ then gives
\begin{align} \label{EeB}
\!\vect E\,\vect e\vect B-\vect B\,\vect e\,\vect E=-2I\vect B\cdot(\vect e\times\vect E)=\nonumber\\
-2I\vect e\cdot(\vect E\times\vect B).
\end{align}

We now evaluate the inner product
\begin{equation}\label{Scprod}
G\cdot\gamma_i=\Big(\frac{E_i}{c}+I\vect B\cdot \vect e_i\Big)\gamma_0=
\Big(\frac{E_i}{c}+\vect e_i\times\vect B\Big)\gamma_0,
\end{equation}
bearing in mind that $\gamma_i=\vect e_i\gamma_0$ and that, in the
three-dimensional relative algebra, $I$ commutes with spatial
vectors:
\begin{equation*}
(I\vect B)\cdot\vect e_i=\frac I2\left(\vect B\vect e_i-\vect e_i\vect B\right)=I(\vect B\wedge\vect e_i)=\vect e_i\times\vect B.
\end{equation*}

\subsection{Geometric Meaning of the Decomposition}

The geometric product of the differential operator and a vector
yields the multivector
\[
\partial A=\langle \partial A \rangle_0+\langle \partial A \rangle_2,
\]
where the scalar part corresponds to the divergence and the bivector
part to the curl.

Thus, the Clifford algebra ($Cl_{1,3}$), constructed on the basis
$\gamma_\mu$, provides a unified formalism.

\end{appendices}

\end{document}